\documentclass[aps,prd,twocolumn,groupedaddress,showpacs,showkeys]{revtex4-1}
\usepackage{graphicx}
\usepackage{dcolumn}
\usepackage{bm}
\usepackage{amssymb,amsmath,latexsym,amsfonts}
\usepackage{float} 
\usepackage{footnote}
\usepackage{graphicx}
\usepackage{dcolumn}
\usepackage{bm}
\usepackage{amssymb,amsmath,latexsym,amsfonts}
\usepackage{float} 
\usepackage{footnote}
\usepackage[margin=0.5in]{geometry}
\usepackage{textcomp}

\usepackage{pgfplots}
\pgfplotsset{width=8cm,compat=1.5}
\begin{document}


\title{Mass bounds for Newtonian bosonic stars in theory of ultra--cold gases}

\author{S. Guti\'errez}
 \email{sergiogs@xanum.uam.mx} \affiliation{Departamento de F\'{\i}sica,
 Universidad Aut\'onoma Metropolitana--Iztapalapa\\
 Apartado Postal 55--534, C.P. 09340, M\'exico, D.F., M\'exico.}

\author{B. Carvente}
 \email{carvente@xanum.uam.mx} \affiliation{Departamento de F\'{\i}sica,
 Universidad Aut\'onoma Metropolitana--Iztapalapa\\
 Apartado Postal 55--534, C.P. 09340, M\'exico, D.F., M\'exico.}

\author{A. Camacho}
 \email{acq@xanum.uam.mx} \affiliation{Departamento de F\'{\i}sica,
 Universidad Aut\'onoma Metropolitana--Iztapalapa\\
 Apartado Postal 55--534, C.P. 09340, M\'exico, D.F., M\'exico.}


 \date{\today}

\begin{abstract} In the present work we address the issue concerning the
relation mass versus radius for bosonic stars, all this in the
context of the theory of ultra--cold gases in the dilute regime and
within the realm of Newtonian gravity. It will be shown that this
model predicts a minimum and a maximum for the mass of this kind of
stars, contrary to the current belief present in some results in
this direction.

\end{abstract}

\pacs{95.35.+d, 95.30.Sf}
\maketitle

\date{\today}

\section{\bf Introduction}
The concept of bosonic star (BS) emerged at the end of the 1960's
\cite{Kaup, Ruffini} and for some time its interest was mainly
related to the understanding of the laws behind its behavior,
without any relation to astronomical or astrophysical observations.
Concerning the possible cases that BS could have let us mention that
the structure of the self--interaction of the basic constituents is
one of the elements (spacetime symmetry has also to be included)
defining the type of BS \cite{Mielke}, charged, rotating, etc. An
interesting point in these models is related to the existence of an
upper bound for the mass, this value defines a region below the one
the BS cannot collapse to a black hole.

Ultra--cold bosonic gases have been considered a possibility for the
description of dark matter \cite{Bohmer} and also for BS
\cite{Chavanis, Chavanis1}, two situations which, without
exaggerating, can be considered theoretical siblings. In this
context the current description for these systems accepts the
possibility of an infinite mass enclosed in a finite region, see
expression (95) in \cite{Chavanis}. A fleeting glimpse at the
results in other approaches allows us to find that they do include
the presence of an upper mass bound for BS \cite{Liebling}. At this
point we may wonder if this discrepancy is a consequence of the fact
that we have different approaches, or something else lies behind it.
In the
 present work, which analyzes BS in the context of ultra--cold gases
 in the dilute regime, we provide an analytical answer in which we
 always bear in mind the mathematical conditions defining the dilute
 requirement. It will be proved that there is, not only, an upper
 limit, but also a lower one. As expected, these two limits depend upon the
 mass and scattering length of the involved bosonic particle.
 These two parameters emerge from the
 fundamental physical restrictions behind the present model in which
 the volume of the star, the number of its constituents, and the
 scattering length, related to the pairwise self--interaction, fulfill an
 inequality which codifies the mathematical premises defining the corresponding
 model.

 In order to have a profounder physical motivation for the present
work we proceed to provide a simple argument (based upon the
fundamentals of an ultra--cold bosonic gas in the dilute regime)
that could shed some light upon this topic, namely, there has to be
an upper bound for the mass of a BS. Indeed, we now assume the
validity of expression (95) in \cite{Chavanis} for a BS, namely, for
a finite radius, here denoted by $R_{(min)}$, the system does not
have an upper bound for its mass, i.e., it may grow indefinitely.
The mass of this star is here denoted by M and it can be written as
the product of the mass of our single particles ($m$) times the
total number of particles ($N$), i.e., $M=mN$. It has to be strongly
underlined that our theory requires the fulfillment of the dilute
gas condition, which is a backbone of the model under consideration
\cite{Pitaevskii1}. This requirement can be cast in the following
form $(V/N)^{1/3}>a$, where $V$ is the volume of the system and $a$
the scattering length of the involved particles which is responsible
for the description of the pairwise interactions taking place in the
gas. Clearly, for a non--vanishing scattering length if $N$ grows
without any bound we reach a value for it, say $N^{(max)}$, such
that $(V/N^{(max)})^{1/3}=a$, for $N^{(+)}>N^{(max)}\Rightarrow
(V/N^{(+)})^{1/3}<a$. In other words, the assertion that a theory
based upon a Gross--Pitaevskii--Poisson equation predicts the
possibility of stars with unbounded mass implies that we have
violated one of the fundamental premises of the current theory,
i.e., the requirement of dilute gas regime. In other words, this
last argument shows that, within the mathematical conditions
defining our model, the number of particles in a bosonic star has an
upper bound. Therefore we may ask ourselves the following question:
how can we deduce this bound for the mass of the BS? The answer to
this question is the topic of the present work.

We have $N$ bosonic particles, each one of them with mass $m$ and
scattering length $a$, defining a spherical body with radius $R$. We
assume that the temperature of this system is smaller than that
corresponding to the beginning of the condensation process. Under
these conditions the description of this system lies within the
context of dilute ultra--cold gases \cite{Pethickbook}, a fact that
implies that our calculations shall always bear in mind the
fulfillment of all the conditions defining the model. Another
element is related to the introduction of gravity as a fundamental
point in the physics of these systems. The set of corresponding
equations are called Gross--Pitaevskii--Poisson \cite{Chavanis}.

\section{\bf Mathematical modeling of the gravitational interaction}

We interpret now the gravitational interaction as a trapping
potential with the structure of an isotropic harmonic oscillator.
Consider a spherical body of mass $M$ and radius $R$ the one has a
small cavity along the diameter coincident with the $z$--axis. The
corresponding mass density function has spherical symmetry. This
mass density function has to be depicted by a function such that its
Taylor expansion about the center of the body renders a series,
i.e., it cannot be a polynomial. Indeed, assume that

\begin{eqnarray}
\rho(r)= \rho(0)
+\sum_{n=1}^{\infty}\rho^{(n)}(0)r^n/n!.\label{equa01}
\end{eqnarray}

Such that $\exists~s\in \mathbb{N}\backepsilon \rho^{(n)}(0)=0,
\forall~n\geq s$. This condition implies that around the center of
the body the density is a polynomial. Therefore

\begin{eqnarray}
r\rightarrow\infty\Rightarrow\lim\rho(r)\rightarrow\infty.\label{equa02}
\end{eqnarray}

Clearly, this condition implies that the density cannot be depicted
by a polynomial. In other words, $\forall~l\in\mathbb{N},
\exists~s\in\mathbb{N}$ such that $\rho^{(s)}(0)\not=0, s\geq l$.

Let us consider one condition upon $\rho(r)$, namely, its global
maximum lies at $r=0$. A particle of mass $m$ moves along this
cavity and the coordinate system has its origin at the geometrical
center of the body. In spherical coordinates

\begin{eqnarray}
m\frac{d^2r}{dt^2}=-G\frac{mM(r)}{r^2}.\label{equa1}
\end{eqnarray}

$M(r)$ defines the mass inside the sphere of radius $r$.

\begin{eqnarray}
M(r)=\frac{4\pi}{3}\rho(0)r^3\Bigl[1+3\sum_{n=2}^{\infty}\frac{\rho^{(n)}(0)}{\rho(0)n!(n+3)}r^n\Bigr].\label{equa2}
\end{eqnarray}

Finally, the equation of motion is

\begin{eqnarray}
\frac{d^2r}{dt^2}+\frac{4\pi}{3}G\rho(0)r\Bigl[1+3\sum_{n=2}^{\infty}\frac{\rho^{(n)}(0)}{\rho(0)n!(n+3)}r^n\Bigr]=0.\label{equa3}
\end{eqnarray}

The series in our last expression exists therefore we may define

\begin{eqnarray}
f(r)\equiv\sum_{n=2}^{\infty}\frac{\rho^{(n)}(0)}{\rho(0)n!(n+3)}r^n.\label{equa002}
\end{eqnarray}

We have that $\forall~\delta >0, \exists~l\in\mathbb{N}$ such that

\begin{eqnarray}
\vert
f(r)-\sum_{n=2}^{s}\frac{\rho^{(n)}(0)}{\rho(0)n!(n+3)}r^n\vert<\delta,
~\forall~s\geq l.\label{equa003}
\end{eqnarray}

Hence

\begin{eqnarray}
\vert\sum_{n=s+1}^{\infty}\frac{\rho^{(n)}(0)}{\rho(0)n!(n+3)}r^n\vert<\delta,
~\forall~s\geq l.\label{equa004}
\end{eqnarray}

If $\delta_1=10^{-1}$, then $\exists ~l_1\in\mathbb{N}$ such that

\begin{eqnarray}
\vert\sum_{n=s+1}^{\infty}\frac{\rho^{(n)}(0)}{\rho(0)n!(n+3)}r^n\vert<10^{-1},
~\forall~s\geq l_1.\label{equa005}
\end{eqnarray}

This last expression entails an inequality for our equation of
motion, indeed

\begin{eqnarray}
-\frac{d^2r}{dt^2}\leq\frac{4\pi}{3}G\rho(0)r\Bigl[\frac{13}{10}+3\sum_{n=2}^{l_1}\frac{\rho^{(n)}(0)}{\rho(0)n!(n+3)}r^n\Bigr].\label{equa006}
\end{eqnarray}

We now consider that our density function has the following
characteristic

\begin{eqnarray}
\vert\frac{\rho^{(n)}(0)}{\rho(0)}\vert\sim\frac{1}{R^n}.\label{equa007}
\end{eqnarray}

This feature appears in several non--compact, for
instance, Gaussian, Lorentzian functions and also all the even
eigenfunctions of a harmonic oscillator.

The motion of our test particle takes place within the region
defined by $r\in[0,R]$ then the ensuing equation of motion has the
following approximate expression

\begin{eqnarray}
\frac{d^2r}{dt^2}
+\frac{4\pi}{3}G\rho(0)r\Bigl[\frac{13}{10}+3\sum_{n=2}^{l_1}\bigl(\frac{r}{R}\bigr)^n\Bigr]=0.\label{equa008}
\end{eqnarray}

The factor $13/10$ defines a three dimensional harmonic oscillator
and the additional terms may be understood as perturbations to it,
at least in those cases in which $r<R$. The corresponding frequency
is given by

\begin{eqnarray}
\omega_{(0)}=\sqrt{\frac{52\pi G\rho(0)}{30}}.\label{equa4}
\end{eqnarray}

This expression tells us that the motion of $m$ is, approximately,
related to an isotropic harmonic oscillator whose frequency depends
upon the central density. We now quantize our system and consider
the gravitational effects of $N-1$ particles of mass $m$
($M=(N-1)m$) upon our particle of, also, mass $m$. The corresponding
time--independent Schr\"odinger equation reads

\begin{eqnarray}
E\psi=-\frac{\hbar^2}{2m}\nabla^2\psi +
\frac{m\omega^2_{(0)}}{2}r^2\psi.\label{equa5}
\end{eqnarray}

Since we end up with a three--dimensional harmonic oscillator the
ground state of system is a Gaussian function \cite{Cohenbook1}. In
addition, the Hartree approximation is introduced \cite{Pethickbook}
then all particles are described by a Gaussian wavefunction, i.e.,
all particles are described by one and only one function, namely,

\begin{eqnarray}
\rho(r)=\frac{mN}{\sqrt{\pi^3}l^3}\exp\{-r^2/l^2\},\label{equa56}
\end{eqnarray}

\begin{eqnarray}
l=\sqrt{\frac{\hbar}{m\omega_{(0)}}}.\label{equa5h6}
\end{eqnarray}

\subsection{\bf Mathematical model}

The physically meaningful case involves a non--vanishing scattering
length therefore we introduce now this parameter. According to our
previous analysis the self--gravitational interaction of the star is
considered as an external isotropic three--dimensional harmonic
oscillator, in other words, the corresponding many
body--Hamiltonian, for the situation of a dilute gas, is given by

\begin{eqnarray}
\hat{H}=\sum_{i=1}^N\Bigl[\frac{p_{(i)}^2}{2m}+\frac{m\omega^2_{(0)}}{2}r_{(i)}^2\Bigr]\nonumber\\
+U_{(0)}\sum_{i<j}\delta\bigl(\vec{r}_{(i)}-\vec{r}_{(j)}\bigr).
\label{equa8}
\end{eqnarray}

Here $U_{(0)}=4\pi\hbar^2a/m$ contains the information, under the
assumption of very low energies and dilute gas, of the interaction
between two particles, i.e., only pairwise interactions are relevant
for the dynamics of the system \cite{Pethickbook}. Here $a$ denotes
the scattering length of our particles.

In the context of the Hartree approximation the time--independent
Gross Pitaevskii equation provides the dynamics of the system
\cite{Uedabook}

\begin{eqnarray}
\mu\psi(\vec{r})= \Bigl[-\frac{\hbar^2}{2m}\nabla^2 +
\frac{m\omega^2_{(0)}}{2}r^2 +
U_{(0)}\vert\psi(\vec{r})\vert^2\Bigr]\psi(\vec{r}).
 \label{equa9}
\end{eqnarray}

An additional condition has to be satisfied, the one implies
particle conservation, namely

\begin{eqnarray}
N= \int\vert\psi(\vec{r})\vert^2d\vec{r}.
 \label{equa10}
\end{eqnarray}

The chemical potential is $\mu$. Since
$\rho(\vec{r})=m\vert\psi(\vec{r})\vert^2$ we may cast (\ref{equa9})
in the following form

\begin{eqnarray}
\mu\psi(\vec{r})= \Bigl[-\frac{\hbar^2}{2m}\nabla^2 +
\frac{m\omega^2_{(0)}}{2}r^2 +
U_{(0)}\rho(\vec{r})/m\Bigr]\psi(\vec{r}).
 \label{equa11}
\end{eqnarray}

Clearly,we have that

\begin{eqnarray}
\vert\psi(\vec{r})\vert^2\psi(\vec{r})=\rho(\vec{r})\psi(\vec{r})/m.
\label{equa12}
\end{eqnarray}

Introducing the Gaussian structure for the density, expanding it in
terms of a Taylor series, and, finally, keeping only terms up to
second order in $r$ we end up with the following expression

\begin{eqnarray}
\tilde{\mu}\psi(\vec{r})= -\frac{\hbar^2}{2m}\nabla^2\psi(\vec{r}) +
\frac{m\omega^2}{2}r^2\psi(\vec{r}), \label{equa13}
\end{eqnarray}

\begin{eqnarray}
\omega^2=\omega^2_{(0)}-2\frac{NU_{(0)}}{m\sqrt{\pi^3}l^5},
\label{equa14}
\end{eqnarray}

\begin{eqnarray}
\tilde{\mu}=\mu-\frac{NU_{(0)}}{\sqrt{\pi^3}l^3}. \label{equa1c4}
\end{eqnarray}

The effective chemical potential is denoted by $\tilde{\mu}$.
Clearly, the equation of motion is a three--dimensional harmonic
oscillator in which the frequency is now modified due to the
presence of a non--vanishing scattering length.

\section{\bf Bounds for the mass parameter}

\subsection{\bf Mechanical Equilibrium}

At this point we address the issue concerning the condition of
mechanical equilibrium for the BEC. Indeed, gravity tends to
collapse the star and this behavior faces a pressure which is a
consequence of Heisenberg's Uncertainty Principle and of the
movement of the particles of the star. Mechanical equilibrium
emerges when the corresponding pressures of these two processes are
equal.

The energy of the system due to $N$ particles in the ground state
(no particles in excited states, as a first approximation) is given
by

\begin{eqnarray}
E_{(0)}=\frac{3}{2}\hbar\omega N. \label{equa15}
\end{eqnarray}

In this last expression the frequency corresponds to (\ref{equa14})
and it does not neglect the kinetic energy, i.e., our formalism does
not resort to the Thomas--Fermi approximation (TF)
\cite{Pethickbook}. The use of TF is valid if $Na>l$, a fact that
requires the knowledge of the parameters $N, a, m$. Clearly, we do
not know them, on the contrary, one of our problems is the deduction
of them; in other words, we cannot resort to TF since we know
nothing about the require parameters of the bosonic particles.

Our system is equivalent to a BEC trapped by a three--dimensional
isotropic harmonic oscillator \cite{Stringaribook1}, a situation
already well comprehended. The pressure related to this case reads:

\begin{eqnarray}
P_{(c)}= \frac{3\alpha^2\hbar^2N}{4\pi mR^5}+\frac{3\alpha^3 U_{(0)}N^2}{4\pi^{5/2}
R^6}. \label{equa21}
\end{eqnarray}

In this last expression $R =\alpha l$ is the characteristic radius
of the region comprising most (at least 87 percent) of the
particles. 

We now address the issue concerning the gravitational attraction of
our bosonic star, a topic related to the equilibrium of a spherical
body with density, pressure, velocity field, and gravitational
potential $\rho$, $P$, and $\vec{v}$, $\Phi$, respectively. The
corresponding equations for the internal structure are provided by
\cite{Willbook}

\begin{eqnarray}
\rho\frac{d\vec{v}}{dt}= \rho\nabla\Phi-\nabla P, \label{equa22}
\end{eqnarray}

\begin{eqnarray}
\frac{\partial\rho}{\partial t}
+\nabla\cdot(\rho\vec{v})=0.\label{equa23}
\end{eqnarray}

These expressions do not close the system, namely, an additional
piece of information is required, i.e., the equation of state; in
other words, the functional dependence among pressure, density,
temperature, etc. The Newtonian gravitational potential for a
spherical body of radius $R$ is

\begin{eqnarray}
\nabla\Phi(t,r)=\frac{GM(t,r)}{r}+4\pi\int_r^R\rho(t,r')dr'.
\label{equa24}
\end{eqnarray}

In this last expression $M(t,r)$ is the mass contained in a sphere
(coincident with our body) of radius $r$. We now consider the
surface of our sphere and calculate the change in this gravitational
potential due to a change in the volume, a process that entails a
pressure ($P_{(g)}=-\frac{\partial\Phi}{\partial V}$).

\begin{eqnarray}
P_{(g)}=\frac{GM^2}{4\pi R^4}.\label{equa25}
\end{eqnarray}

The corresponding pressure is a non--constant function of the radial
distance \cite{Willbook} and, in consequence, we must identify the
value of $r$ related to (\ref{equa25}). This pressure is deduced
after evaluating the gravitational potential on the surface of the
body then it represents the pressure on this surface. The
mathematical condition determining mechanical equilibrium is the
equality of our two pressures on the surface of the bosonic star
(remember that $M=mN$) i.e., expressions (\ref{equa21}) and
(\ref{equa25}).

One consequence of the equality $P_{(g)}=P_{(c)}$ is related to the
fact that we have deduced $R$ as a function of $m$, $a$, and $N$. In
this sense, the roughest approximation yields the following
expression for the radius of the bosonic star.

    \begin{equation}\label{radio}   R=\frac{3}{2}\frac{\alpha^2\hbar^2}{Gm^3N}\left\{1+\sqrt{1+\frac{16aN^2Gm^3}{3\alpha \sqrt{\pi}\hbar^2}}\right\}.
    \end{equation}

    Defining  $L\equiv\frac{\alpha^2\hbar^2}{Gm^3 a}$ and inserting the dilute gas condition into (\ref{radio}), $\left(\frac{V}{N}\right)^{1/3}>a$,
    we have an inequality involving the number of particles within the boson star:
    \begin{equation}\label{inequality}
    \beta\equiv\sqrt{1+\frac{16\alpha N^2}{3\sqrt{\pi}L}}>\frac{2}{3}\left(\frac{3}{4\pi}\right)^{1/3}\frac{N^{4/3}}{L}-1\equiv\gamma.
    \end{equation}

In this last expression, we have two possibilities about $\gamma$
and $\beta$:
\begin{enumerate}

    \item $\beta>\gamma>0$,
    \item $\gamma<0 \wedge |\beta|<|\gamma|$.
\end{enumerate}

The second case leads us to the following conclusion:

$1-\frac{(0.41)N^{4/3}}{(1.67)^2L}>\sqrt{1+\frac{5N^2}{l}}>0$, i.e.,
a condition impossible to satisfy. In other words, the only
mathematically consistent situation is the first one.

Therefore, concerning the only possibility we now define a
polynomial in the variable as follows $z=N^{2/3}$:
\begin{equation}\label{eqPolinom}
P(z)=\frac{z^2}{L}\left\{\frac{4}{9L}\left(\frac{3}{4\pi}\right)^{2/3}z^2-\frac{16\alpha}{3\sqrt{\pi}}z-\frac{4}{3}\left(\frac{3}{4\pi}\right)^{1/3}\right\},
\end{equation}

and solving for $z$, the only physical solution reads
$$z_{(+)}= \left(\frac{4\pi}{3}\right)^{2/3}\frac{6\alpha L}{\sqrt{\pi}}.$$ Relating this last expression to the original
variable $N$ we have a maximum value for the total mass in a bosonic
star:
\begin{equation}\label{maxMass}
M_{(+)}=mN_{(+)}=mz_{(+)}^{3/2}=
m\frac{4\pi}{3}\left(\frac{6}{\sqrt{\pi}}\right)^{3/2}(\alpha L)^{3/2}.
\end{equation}
In addition we have the following condition
$\frac{2}{3}\left(\frac{3}{4\pi}\right)^{1/3}\frac{N^{4/3}}{L}>1$,
it implies the existence of a lower bound for the mass of a bosonic
star:

\begin{equation}\label{minMass}
M_{(1)}=m\left[\frac{3}{2}\left(\frac{4\pi}{3}\right)^{1/3}L\right]^{4/3}
.\end{equation} Resorting to equation (\ref{radio}) we may estimate,
in a rough approximation, the parameter $a/m^3$ for a given radius:
\begin{equation}\label{parameter}
\frac{a}{m^3}=\frac{1}{12}\frac{G\sqrt{\pi}}{\alpha^3\hbar^2}R^2.
\end{equation}

Equation (\ref{parameter}) and the behavior of our bosonic particles
in the context of dilute gases \cite{Gutierrez} for the parameter
$a/m^3$ allow us to compare and predict the value of the mass
particle for bosonic stars from the typical size of stars made of
baryonic matter.
\\ \\

\section{DISCUSSION OF RESULTS AND CONCLUSIONS}

We now proceed to analyze the deduction of the values of the mass
and scattering length of our bosonic particles. In order to do this
we now resort to expression (\ref{parameter}) which is function of
$R$. Notice that the choice of a value for $R$ defines, uniquely,
the value of $a/m^3$. At this point we notice that for baryonic
particles (this phrase means Na, K, and Rb atoms) and for the
results associated to dark matter halos \cite{Gutierrez} the graph
$a/m^3$ versus $m$ in a logarithmic version renders a straight line.
We now introduce an additional assumption, namely, the relation
$a/m^3$ as a function of $m$ is a fundamental expression for
baryonic and exotic matter. Therefore, the choice of a value for $R$
provides us a unique $a/m^3$ and, resorting to the next plot, we
deduce $m$, a fact that implies also the knowledge of $a$.

At this point we may use the method of least squares to provide a
curve to determine the corresponding values for $m$ for the
bosonic star particles. The associated plot is\\

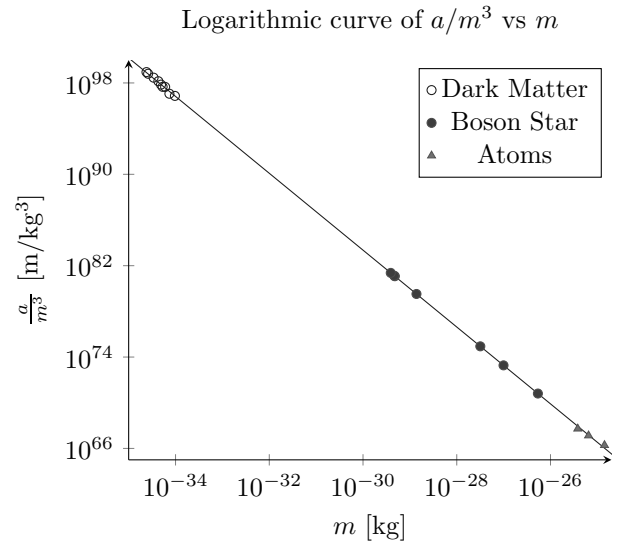
\nocite{*}
\selectcolormodel{gray}
\begin{tikzpicture}
\begin{loglogaxis}[
    title={Logarithmic curve of $a/m^3$ vs $m$},
    axis lines=left,
    xlabel={$m$ [kg]},
    ylabel={$\frac{a}{m^3}$ [m/kg$^3$]},
    ymin=10^65, ymax=10^100,
    xmin=10^-35, xmax=2*10^-25,]
\addplot+[
    only marks,
    mark=o,
    mark size=1.7pt,]
    coordinates {
    (2.4e-35,8.89757e+98)(2.6e-35,6.37233e+98)(3.4e-35,2.90047e+98)(4.3e-35,1.40868e+98)(4.9e-35,7.13988e+97)(5.3e-35,4.56753e+97)(6e-35,4.49074e+97)(7.4e-35,1.13517e+97)(9.7e-35,7.34107e+96)};
\addplot+[
    only marks,
    mark=*,
    mark size=1.7pt,]
    coordinates {
    (5.38654e-27,6.382e+70)(9.94871e-28,1.876e+73)(3.18254e-28,8.686e+74)(1.3816e-29,3.335e+79)(4.76946e-30,1.195e+81)(4.64838e-30,1.303e+81)(3.90802e-30,2.336e+81)};
\addplot+[only marks,
   mark=triangle*,
   mark size=1.7pt,]
   coordinates {
   (3.81754e-26,5.26761e+67)(6.49243e-26,1.33378e+67)(1.41923e-25,1.90606e+66)};
\addplot[
    blue,
    domain=10^-38:10^-24,]
    {10^(-17.5858) * x^(-3.36487)};
   \legend{Dark Matter,Boson Star,Atoms}
\end{loglogaxis}
\end{tikzpicture}

Concerning the main expression (\ref{parameter}), we may establish
the value for $a$ associated to the bosonic star particle.
In addition, now we can provide an explicit value for the mass bounds given by (\ref{minMass}) and (\ref{maxMass}), and these are:\\

\begingroup
\renewcommand{\arraystretch}{1.3}
\begin{table}[H]
    \centering
\begin{tabular}{l c c c c}
    Star & Radius $R$  & $a/m^3$ & $m$ & $a$ \\
    & [km] & [m/kg$^3$] & [MeV/c$^2$] & [m] \\
    \hline
    \hline
    White dwarf & $6.0$ x $10^{3}$ & $6.9$ x $10^{69}$ & $5.9$ x $10^{3}$ & $7.9$ x $10^{-9}$ \\
    Brown dwarf & $4.2$ x $10^{4}$ & $3.4$ x $10^{71}$ & $1.8$ x $10^{3}$ & $1.2$ x $10^{-8}$ \\
    Sun & $7.0$ x $10^{5}$ & $9.3$ x $10^{73}$ & $3.5$ x $10^{2}$ & $2.2$ x $10^{-8}$ \\
    Red giant  & $5.6$ x $10^{7}$ & $6.0$ x $10^{77}$ & $26.0$ & $5.7$ x $10^{-8}$ \\
    Supergiant & $3.5$ x $10^{8}$ & $2.3$ x $10^{79}$ & $8.6$ & $8.5$ x $10^{-8}$ \\
    Betelgeuse & $8.2$ x $10^{8}$ & $1.3$ x $10^{80}$ & $5.2$ & $1.0$ x $10^{-7}$ \\
    NML Cygni & $1.1$ x $10^{9}$ & $2.5$ x $10^{80}$ & $2.3$ & $1.1$ x $10^{-7}$\\
    \hline
\end{tabular}
    \end{table}

    \renewcommand{\arraystretch}{1.3}
    \begin{table}[H]
        \centering
        \begin{tabular}{l c c c }
            Star & Radius $R$ & $M_{(1)}$ & $M_{(+)}$\\
            & [m] & [kg] & [kg] \\
            \hline
            \hline
            White dwarf & $6.0$ x $10^{3}$ & $3.3$ x $10^{-18}$ & $3.5$ x $10^{-12}$ \\
            Brown dwarf & $4.2$ x $10^{4}$ & $6.0$ x $10^{-17}$ & $1.1$ x $10^{-10}$ \\
            Sun & $7.0$ x $10^{5}$ & $4.0$ x $10^{-15}$ & $1.5$ x $10^{-8}$ \\
            Red giant  & $5.6$ x $10^{7}$ & $2.8$ x $10^{-12}$ & $3.2$ x $10^{-5}$ \\
            Betelgeuse & $8.2$ x $10^{8}$ & $4.3$ x $10^{-11}$ & $8.0$ x $10^{-4}$\\
            Supergiant & $3.5$ x $10^{8}$ & $1.6$ x $10^{-10}$ & $3.6$ x $10^{-3}$\\
            NML Cygni & $1.1$ x $10^{9}$ & $2.6$ x $10^{-10}$ & $6.5$ x $10^{-3}$ \\
            \hline
        \end{tabular}
    \end{table}

Our previous results show us that the theory of ultra--cold gases
predicts that the relation between radius and the parameter $a/m^3$
is a unique one. Indeed, notice that stars with different radii
require exotic particles with different mass, and that they may show
a discrepancy of three orders of magnitude among them. Notice that
the values of $a$ and $m$ are far from those corresponding to a dark
matter particle \cite{Gutierrez}, at least in five orders for the
biggest stars like Betelgeuse. An evaluation of the temperature
($T$) of the system can be obtained resorting to the equipartion
theorem \cite{Cohenbook1}, i.e., $3\hbar\omega/2=\kappa T$.

We may accept, at this point, the possible situation in which there
are more than one different kind of exotic particles. This
assumption opens up the possibility of having several exotic
particles, one responsible for a dark matter halo, additional
options defining bosonic stars, with different radii. If we assume,
instead, that only one kind of non-baryonic particle exists, then
there is one and only one possible structure conformed by exotic
particles.

Summing up, we have modeled bosonic stars as a Bose-Einstein
condensate and interpreted the self-gravitational interaction as an
isotropic three-dimensional harmonic oscillator, of course, the
pairwise interaction of short range, emerging in the context of the
dilute limit, has also been considered. We have deduced upper and
lower bounds for the mass of these kind of objects, which appear as
consequence of the mathematical conditions defining our model. The
deduced values show a strong dependence in the choice of the radius
of the star, a fact that implies that if we accept the possibility
of having bosonic stars with different radii, then the present work
tells us that there have to be several exotic particles, i.e., one
of them cannot account for the difference in the radii.

Finally, if we accept the fact that there is only one kind of exotic
particle and that it is related to a dark matter halo
\cite{Gutierrez}, then our results imply that a dilute
Bose--Einstein condensate prevents the appearance subgalatic
structures. On the other hand, if we accept the existence of
subgalatic structures, then galaxies cannot be explained.

\begin{acknowledgements}
B. C. acknowledges CONACyT grant No. 574365 and S. G. the received
UAM grant.
\end{acknowledgements}

\end{document}